\documentclass[prd,aps,eqsecnum,floatfix,nofootinbib,preprint,tightenlines]{revtex4}

\usepackage{latexsym}
\usepackage{graphicx}
\usepackage{multirow}
\usepackage[dvipsnames]{xcolor}

\let\inodot=\i

\def\i{\iota}

\def\cbo{{\,\raise-.15ex\Sc [\,}}                       

\def\ddt#1{{\buildrel {\hbox{\LARGE .\kern-2pt.}} \over {#1}}}

\def\ttl#1{{\it #1}}

\begin{document}

\begin{boldmath}
\begin{center}
{\large{\bf
Data-based determination of the isospin-limit light-quark-connected
contribution to the anomalous magnetic moment of the muon}
}\\[8mm]
Diogo Boito,$^a$ Maarten Golterman,$^{b,c}$ Kim Maltman$^{d,e}$ and 
Santiago Peris$^c$\\[8 mm]
$^a$Instituto de F{\'\inodot}sica de S{\~a}o Carlos, Universidade de S{\~a}o 
Paulo\\ CP 369, 13570-970, S{\~a}o Carlos, SP, Brazil
\\[5mm]
$^b$Department of Physics and Astronomy, San Francisco State University,\\
San Francisco, CA 94132, USA
\\[5mm]
$^c$Department of Physics and IFAE-BIST, Universitat Aut\`onoma de Barcelona\\
E-08193 Bellaterra, Barcelona, Spain
\\
[5mm]
$^d$Department of Mathematics and Statistics,
York University\\  Toronto, ON Canada M3J~1P3
\\[5mm]
$^e$CSSM, University of Adelaide, Adelaide, SA~5005 Australia
\\[10mm]
\end{center}
\end{boldmath}
\begin{quotation}
We describe how recent determinations of exclusive-mode contributions 
to $a_\mu^{\rm LO,HVP}$, the leading-order hadronic vacuum polarization 
contribution to the anomalous magnetic moment of the muon, can 
be used to provide, up to small electromagnetic (EM) corrections 
accessible from the lattice, a data-based dispersive determination 
of $a_\mu^{\rm lqc;\, IL}$, the isospin-limit, light-quark-connected 
contribution to $a_\mu^{\rm LO,HVP}$. Such a determination is of 
interest in view of the existence of a number of lattice results for 
this quantity, emerging evidence for a tension between lattice and 
dispersive determinations of $a_\mu^{\rm LO,HVP}$, and the desire 
to clarify the source of this tension. Taking as input for the small 
EM correction that must be applied to the purely data-driven dispersive
determination the result $-0.93(58)\times 10^{-10}$ obtained in 
a recent BMW lattice study, we find $a_\mu^{\rm lqc;\, IL}$ to be 
$635.0(2.7)\times 10^{-10}$ if the results of Keshavarzi, Nomura 
and Teubner are used for the exclusive-mode contributions and 
$638.4(4.1)\times 10^{-10}$ if instead those of Davier, H\"ocker, 
Malaescu and Zhang are used.
\end{quotation}

\newpage
\section{Introduction}\label{intro}
As is well known, the recent FNAL E989 result~\cite{FNL} for 
the anomalous magnetic moment of the muon, $a_\mu =(g-2)/2$, is 
in good agreement with the earlier BNL E821 result~\cite{BNL},
producing an updated world average which is $4.2\sigma$ larger than
the Standard Model (SM) expectation detailed in the $g-2$ Theory 
Initiative assessment~\cite{review}, and based on the work of
Refs.~\cite{Aoyama:2012wk,Aoyama:2019ryr,Czarnecki:2002nt,Gnendiger:2013pva,Davier:2017zfy,Keshavarzi:2018mgv,Davier:2019can,KNT19,Colangelo:2018mtw,Hoferichter:2019mqg,Hoid:2020xjs,Kurz:2014wya,Melnikov:2003xd,Masjuan:2017tvw,Colangelo:2017qdm,Colangelo:2017fiz,Hoferichter:2018dmo,Hoferichter:2018kwz,Gerardin:2019vio,Bijnens:2019ghy,Colangelo:2019lpu,Colangelo:2019uex,Colangelo:2014qya,Blum:2019ugy}.

In determining SM contributions to $a_\mu$, a particularly important 
role is played by the leading-order, hadronic vacuum polarization 
contribution, $a_\mu^{\rm LO,HVP}$, the uncertainty on which dominates that 
on the SM expectation for $a_\mu$. The assessment of $a_\mu^{\rm LO,HVP}$ 
arrived at in Ref.~\cite{review} is based on the results of two 
analyses~\cite{Davier:2019can,KNT19} using the standard dispersive 
representation~\cite{Brodsky:1967sr,Lautrup:1968tdb,Gourdin:1969dm}
with current $e^+ e^- \rightarrow {\rm hadrons}$ cross-section data as 
input, supplemented by additional input from 
Refs.~\cite{Colangelo:2018mtw,Hoferichter:2019mqg} for $\pi\pi$ 
contributions below $1\ {\rm GeV}^2$ and $3\pi$ contributions.
The SM expectation for $a_\mu^{\rm LO,HVP}$ can, however, also be
obtained on the lattice using the alternate, weighted Euclidean integral 
representation~\cite{Lautrup:1971jf,deRafael:1993za,Blum:2002ii} and
there has been intense recent activity in the lattice community aimed at 
reducing lattice errors to a level sufficient to make lattice results 
competitive with the dispersive determination~\cite{Chakraborty:2014mwa,RBC:2015you,Chakraborty:2016mwy,Budapest-Marseille-Wuppertal:2017okr,RBCUKQCD:2016clu,Giusti:2017jof,RBC:2018dos,Giusti:2018mdh,Giusti:2018vrc,Gulpers:2018mim,Giusti:2019xct,Shintani:2019wai,FermilabLattice:2019ugu,Gerardin:2019rua,Aubin:2019usy,Giusti:2019hkz,Borsanyi:2020mff,Lehner:2020crt,Aubin:2021vej,Giusti:2021dvd,Risch:2021hty,Wang:2022lkq,Aubin:2022hgm,Ce:2022kxy,Alexandrou:2022amy,FermilabLattice:2022izv,RBC22,FHM22}. 
The most recent lattice result from the BMW 
collaboration~\cite{Borsanyi:2020mff} comes close to this goal, reaching, 
for the first time, sub-percent precision. This result, however, is
in tension with the dispersive value for $a_\mu^{\rm LO, HVP}$.
This tension is also seen, at an enhanced (up to $3.9\sigma$) level, in 
comparisons of dispersive~\cite{Borsanyi:2020mff,Colangelo:2022vok}
and lattice~\cite{Aubin:2019usy,Borsanyi:2020mff,Lehner:2020crt,Wang:2022lkq,Aubin:2022hgm,Ce:2022kxy,Alexandrou:2022amy,RBC22,FHM22}
results for the intermediate window quantity, $a_\mu^W$, introduced by the 
RBC/UKQCD collaboration~\cite{RBC:2018dos}. Similar tensions, reaching
as much as $3.7\sigma$, are also found for the one-sided windows considered 
in Ref.~\cite{FermilabLattice:2022izv}. 

Lattice results for $a_\mu^{\rm LO, HVP}$ are typically quoted as sums of 
isospin-limit light-, strange-, charm- and bottom-quark connected 
contributions, the disconnected contribution, and strong-isospin-breaking
(SIB) and electromagnetic (EM) contributions, the latter two receiving 
both connected and disconnected contributions. By far the largest of 
these contributions is the isospin-limit light-quark-connected one, 
denoted $a_\mu^{\rm lqc;\, IL}$ in what follows. Although 
continuum/dispersive estimates exist for the SIB 
contribution~\cite{James:2021sor,Colangelo:2022prz}, and, from our recent
work~\cite{Boito:2022rkw}, for the sum of the strange-quark-connected and 
full disconnected contributions, no dispersive result exists for 
$a_\mu^{\rm lqc;\, IL}$. In this paper, we remedy this situation, 
performing an almost-completely data-based determination of 
$a_\mu^{\rm lqc;\, IL}$. Our result confirms the existence of a 
tension between the dispersive determination and the most precise 
of the lattice determinations~\cite{Borsanyi:2020mff}. We also 
indicate how the same analysis strategy can, in principle, be employed to 
obtain isospin-limit, light-quark-connected contributions to analogous 
reweighted/window observables such as the RBC/UKQCD intermediate 
window quantity, $a_\mu^{\rm W}$~\cite{RBC:2018dos}, the one-sided window
quantities of Ref.~\cite{FermilabLattice:2022izv}, and the
window quantities of Ref.~\cite{Boito:2022njs}, based on $s$-dependent 
weights designed to produce results for the associated dispersive 
determinations focused on more limited regions in $s$. Work on
the reweighted exclusive-mode integrals needed to complete such analyses
is in progress, and will be reported on elsewhere. In the present paper 
we focus on the quantity $a_\mu^{\rm lqc;\, IL}$, where existing results 
for exclusive-mode contributions to $a_\mu^{\rm LO, HVP}$ already make
a dispersive determination possible.

The rest of the paper is organized as follows. In Section~\ref{ILidea} we 
detail how, in the isospin limit, dispersive results for the exclusive-mode 
contributions to $a_\mu^{\rm LO,HVP}$ can be used to provide a dispersive 
determination of $a_\mu^{\rm lqc;\, IL}$. In Section~\ref{ILimplementation} we 
implement this analysis, ignoring for the moment isospin-breaking (IB)
corrections, using as input for the exclusive-mode contributions those 
determined in Refs.~\cite{Davier:2019can,KNT19}. In Section~\ref{IBcorrns}, 
we discuss (and evaluate) IB corrections to the results obtained in 
Sec.~\ref{ILimplementation}. This section also contains our final results 
for $a_\mu^{\rm lqc;\, IL}$. Finally, in Section~\ref{conclusions} we 
provide a brief summary and discuss how, using more detailed exclusive-mode 
information, in the form of the $s$-dependence of exclusive-mode 
contributions to the R-ratio, $R(s)$, analogous results for the 
isospin-limit, light-quark-connected contributions to differently 
weighted integrals over $R(s)$, such as the window quantities noted 
above, can also be obtained.

\section{Basic analysis strategy}\label{ILidea}
\subsection{Notation}\label{notation}
We begin with some basic notation and associated isospin decompositions.

Key objects in the analysis are the two-point functions, 
$\Pi^{ab}_{\mu\nu}(q)$, of the flavor-octet, vector currents
$V^a_\mu =\bar{q}{\frac{\lambda^a}{2}}\gamma_\mu q$, $a=1,\cdots 8$,
(where $\lambda^a$ are the usual Gell-Mann matrices)
together with the associated polarizations, $\Pi^{ab}(Q^2)$, 
subtracted polarizations, $\hat{\Pi}^{ab}(Q^2)$ and  
spectral functions, $\rho^{ab}(s)$, defined by
\begin{eqnarray}
&&\Pi^{ab}_{\mu\nu}(q) = (q_\mu q_\nu - q^2 g_{\mu\nu})\Pi^{ab}(Q^2) =
i\, \int d^4 x e^{iq\cdot x} \langle 0 \vert 
T \left( V^a_\mu (x) V^b_\nu (0) \right)\vert 0\rangle \, , \label{poln}\\
&&\hat{\Pi}^{ab}(Q^2)=\Pi^{ab}(Q^2)-\Pi^{ab}(0)\, ,\\
&&\rho^{ab}(s)={\frac{1}{\pi}}\, \mbox{Im}\, \Pi^{ab}(Q^2)\ ,\quad
(s=\, -Q^2>0)\ ,\label{specfunc}
\end{eqnarray}
where $s=q^2$ and $Q^2\, =\, -q^2$. 

The decomposition of the $u,\, d,\, s$ part of the EM current, 
$J_\mu^{\rm EM}$, into isovector ($a=3$) and isoscalar ($a=8$) 
parts,
\begin{equation}
J_\mu^{\mathrm{EM}} = V_\mu^3 + {\frac{1}{\sqrt{3}}} V_\mu^8 \,\equiv\,
J_\mu^{\mathrm{EM},3}+J_\mu^{\mathrm{EM},8}\ ,
\label{emisodecomp}\end{equation}
leads to the following decompositions for the subtracted three-flavor EM 
vacuum polarization, $\hat{\Pi}_{\rm EM}(Q^2)$, and spectral function, 
$\rho_{\rm{EM}}(s)$,
\begin{eqnarray}
\hat{\Pi}_{\rm{EM}}(Q^2)&=&\hat{\Pi}_{\rm{EM}}^{33}(Q^2)
+{\frac{2}{\sqrt{3}}}
\hat{\Pi}_{\rm{EM}}^{38}(Q^2)+{\frac{1}{3}}
\hat{\Pi}_{\rm{EM}}^{88}(Q^2)\nonumber\\
&\equiv&  \hat{\Pi}_{\rm{EM}}^{I=1}(Q^2)
+\hat{\Pi}_{\rm{EM}}^{\rm{MI}}(Q^2)+
\hat{\Pi}_{\rm{EM}}^{I=0}(Q^2)\ ,\nonumber\\
\rho_{\rm{EM}}(s)&=&\rho^{33}(s)+{\frac{2}{\sqrt{3}}}\rho^{38}(s)+
{\frac{1}{3}}\rho^{88}(s)\nonumber\\
& \equiv & \rho_{\mathrm{EM}}^{I=1}(s)+
\rho_{\rm{EM}}^{\rm{MI}}(s)+\rho_{\rm{EM}}^{I=0}(s)\, ,
\label{piemrhoemdecomps}\end{eqnarray}
with the $ab=33$ parts pure isovector, the $ab=88$ parts pure isoscalar
and the $ab=38$ parts mixed isospin terms which vanish in the isospin
limit. In the isospin limit $\hat{\Pi}^{33}$ is pure light-quark-connected 
while $\hat{\Pi}^{88}$ contains light-quark-connected, strange-quark-connected 
and all disconnected contributions. 

The leading-order hadronic contribution to $a_\mu$, $a_\mu^{\rm{LO,HVP}}$, is
related to $R(s)=12\pi^2\, \rho_{\rm EM}(s)$ by the standard ``dispersive'' 
representation~\cite{Brodsky:1967sr,Lautrup:1968tdb,Gourdin:1969dm},
\begin{equation}
a_\mu^{\rm{LO,HVP}}={\frac{\alpha_{\rm{EM}}^2m_\mu^2}{9\pi^2}}
\int_{m_\pi^2}^\infty ds\, {\frac{\hat{K}(s)}{s^2}} R(s)\, ,
\label{amudispform}\end{equation}
with $\alpha_{\rm EM}$ the EM fine-structure constant, $R(s)$ determined
from the bare inclusive hadronic electroproduction cross section,
$\sigma^{(0)} [e^+ e^-\rightarrow hadrons(+\gamma)]$, by
\begin{equation}
R(s)={\frac{3s}{4\pi \alpha_{\mathrm{EM}}^2}}\, 
\sigma^{(0)} [e^+ e^-\rightarrow hadrons(+\gamma)]\, ,
\label{rdefn}\end{equation}
and the kernel $\hat{K}(s)$ an exactly known, slowly (and monotonically) 
increasing function of $s$ (see, {\it e.g.}, Ref.~\cite{review}). 
Dispersive determinations of $a_\mu^{\rm LO,HVP}$ are typically obtained by
summing (i) exclusive-mode contributions up to just below $s=4$ GeV$^2$, 
(ii) narrow charm and bottom resonance contributions, and (iii) contributions
evaluated using inclusive $R(s)$ data and/or perturbative QCD (pQCD) in 
the remainder of the high-$s$ region. The exclusive-mode regions in the
analyses of Refs.~\cite{Davier:2019can} and \cite{KNT19} we employ below
are $s\le (1.8\ {\rm GeV})^2 = 3.24\ {\rm GeV}^2$ and
$s\le (1.937\ {\rm GeV})^2 = 3.7520\ {\rm GeV}^2$, respectively.

The isospin decomposition of $\hat{\Pi}_{\rm EM}$ in 
Eqs.~(\ref{piemrhoemdecomps}) leads to the related decomposition for the 
three-flavor ($u,\, d,\, s$) contribution to $a_\mu^{\rm{LO,HVP}}$,
\begin{equation}
a_\mu^{\rm{LO,HVP}} = a_\mu^{33}+{\frac{2}{\sqrt{3}}}a_\mu^{38}
+{\frac{1}{3}} a_\mu^{88}\equiv a_\mu^{I=1}+a_\mu^{\rm{MI}}+a_\mu^{I=0}\, .
\label{amulohvpbreakdown}\end{equation}
To first order in $m_d-m_u$ there are no SIB contributions to either 
$a_\mu^{33}$ or $a_\mu^{88}$, while SIB is expected to dominate $a_\mu^{38}$. 

In what follows, we will also denote contributions from an individual 
exclusive mode, $X$, to $a_\mu^{\rm{LO,HVP}}$, $a_\mu^{I=1}$, $a_\mu^{I=0}$ 
and $a_\mu^{\rm{MI}}$ by $[a_\mu^{\rm{LO,HVP}}]_X$, $[a_\mu^{I=1}]_X$, 
$[a_\mu^{I=0}]_X$ and $[a_\mu^{\rm{MI}}]_X$, respectively. 

\subsection{The basic idea}
The basic idea underlying the analysis is as follows.
In the isospin limit, $\hat{\Pi}_{\rm{EM}}^{I=1}$ is pure
light-quark-connected, while $\hat{\Pi}_{\rm{EM}}^{I=0}$
is a sum of light-quark-connected, strange-quark-connected and all
disconnected contributions, with light-quark-connected contribution 
\begin{equation}
\left[\hat{\Pi}_{\rm{EM}}^{I=0}\right]^{\rm lqc}\, 
=\, {\frac{1}{9}}\,\hat{\Pi}_{\rm{EM}}^{I=1}\ .
\label{i0lqcpihat}\end{equation}
The full light-quark-connected contribution to $\hat{\Pi}_{\rm EM}$
is thus
\begin{equation}
\hat{\Pi}_{\rm{EM}}^{\rm{lqc}}\equiv
{\frac{10}{9}}\, \hat{\Pi}_{\rm{EM}}^{I=1}
\label{lqcpihat}\end{equation}
and the corresponding spectral function
\begin{equation}
\rho_{\rm{EM}}^{\rm{lqc}}(s)={\frac{10}{9}}\,
\rho_{\rm{EM}}^{I=1}(s)\, .
\label{lqcspecfunc}\end{equation}
The desired light-quark-connected contribution to $a_\mu^{\rm LO,HVP}$,
$a_\mu^{\rm lqc;\, IL}$, is then given by the following dispersive integral
involving the $I=1$ spectral function:
\begin{equation}
a_\mu^{\rm lqc,\, IL}={\frac{\alpha_{\rm{EM}}^2m_\mu^2}{9\pi^2}}
\int_{m_\pi^2}^\infty ds\, {\frac{\hat{K}(s)}{s^2}} 
\, \left[ \left({\frac{10}{9}}\right)\, 12\pi^2\, 
\rho_{\rm{EM}}^{I=1}(s)\right]\, .
\label{amulqcdispform}\end{equation}
An accurate determination of $a_\mu^{\rm lqc;\, IL}$ is thus possible 
provided the $I=1$ contribution to $R(s)$, or equivalently 
$\rho_{\rm EM}(s)$, can be identified with sufficient precision. 

The separation of $I=1$ and $I=0$ contributions is straightforward in the 
higher-$s$ inclusive region, where $R(s)$ is approximated using pQCD and
the $I=1$ part represents $3/4$ of the total. In 
the lower-$s$ region, where $R(s)$ is obtained as a sum over exclusive-mode 
contributions, the separation is also straightforward for those exclusive 
modes having well-defined $G$-parity since states with positive/negative 
$G$-parity necessarily have $I=1$/$I=0$. This provides unique isospin 
assignments for contributions from exclusive modes consisting entirely 
of narrow and/or strong-interaction-stable states having well-defined 
$G$-parity ($\pi$, $\eta$, $\omega$, $\phi$), which constitute more than 
$93\%$ of the total exclusive-mode-region contribution. Further input
is needed to separate the $I=1$ and $I=0$ components of contributions 
from exclusive modes which are not eigenstates of $G$-parity, such as those 
containing at least one $K\bar{K}$ pair. We outline in the next section 
how this separation is accomplished using experimental input for the 
$K\bar{K}$ and $K\bar{K}\pi$ exclusive modes. For all other
$G$-parity-ambiguous exclusive modes, $X$, we employ a ``maximally 
conservative'' assessment in which the $I=1$ contribution,
$\left[ a_\mu^{I=1}\right]_X$, is taken to be $50\pm 50\%$ 
of the total $\left[ a_\mu^{\rm{LO,HVP}}\right]_X$. Fortunately, spectral 
contributions from these additional $G$-parity-ambiguous modes lie at 
higher $s$ and thus have contributions to $a_\mu^{\rm{LO,HVP}}$, and
hence also $I=0/1$ separation uncertainties, which are strongly numerically 
suppressed, in spite of their $100\%$ uncertainties.

In Sec.~\ref{ILimplementation} we will implement the above analysis framework
using as input the exclusive-mode results of Refs.~\cite{Davier:2019can,KNT19},
neglecting, to begin with, isospin-breaking (IB) corrections. The 
resulting nominal $a_\mu^{\rm lqc;\, IL}$, which we will denote by
$a_\mu^{\rm lqc}$, will differ from the desired isospin-limit value
by small IB contributions. These IB contributions are taken into 
account and removed in Sec.~\ref{IBcorrns}, which contains our final 
results for $a_\mu^{\rm{lqc;\, IL}}$.

\section{A data-based implementation ignoring 
isospin-breaking effects}\label{ILimplementation}
In this section we carry out two determinations of $a_\mu^{\rm lqc}$,
neglecting IB corrections. These differ in the input used for the 
exclusive-mode $a_\mu^{\rm LO,HVP}$ contributions, one employing
the results of Ref.~\cite{KNT19} (KNT19), the other those of 
Ref.~\cite{Davier:2019can} (DHMZ). The reader is reminded that the 
KNT19 and DHMZ exclusive-mode regions are different, the former 
extending up to $s= 3.7520\ {\rm GeV}^2$, the latter up to only 
$s= 3.24\ {\rm GeV}^2$. Contributions from the region above these 
exclusive-mode endpoints will be obtained using pQCD,{\footnote{The
shorthand ``pQCD'' refers here, and in what follows, to dimension $D=0$, 
mass-independent perturbative OPE contributions. For the $I=1$ 
polarization $\Pi_{\rm EM}^{I=1}$ considered this paper, 
mass-dependent $D=2$ perturbative corrections are 
$O\left(m_{u,d}^2\right)$, and numerically negligible.}}, with an error 
component, to be discussed below, designed to take into account 
possible small duality-violating (DV) contributions. The higher 
onset in $s$ of the inclusive region for KNT19 increases the expected 
accuracy of the pQCD result used to represent $\rho_{\rm EM}^{I=1}(s)$ 
in this region and constitutes an advantage for the determination 
employing KNT19 input over that employing DHMZ input. Details
of the form of the pQCD representation used in the inclusive region,
together with our strategy for estimating possible residual
DV corrections, are provided in Sec.~\ref{inclregionsec}.

The KNT19- and DHMZ-based analyses are outlined in 
Sec.~\ref{knt19dhmzanalyses} below. In both cases we take advantage 
of results already worked out in Ref.~\cite{Boito:2022rkw}. This 
includes results for the data-based $I=0/1$ separation of $K\bar{K}$ 
and $K\bar{K}\pi$ contributions. For the reader's benefit, the 
paragraphs which follow briefly review the treatment of the 
contributions from these modes. Further details may be found in 
Secs.~IV.B and IV.C of Ref.~\cite{Boito:2022rkw}. 

$K\bar{K}$ contributions to $a_\mu^{\rm LO, HVP}$ are expected to be
dominated by the $I=0$ contribution of the $\phi$ resonance, with a much
smaller $I=1$ contribution. This qualitative expectation can be quantified
by combining the electroproduction-based result for the sum of $I=1$ and $0$ 
contributions with recent BaBar results~\cite{BaBarKKbar} for the differential
$\tau^-\rightarrow K^- K^0\nu_\tau$ decay distribution, which provides
an experimental determination of the charged $I=1$ vector current spectral
function, and, via the Conserved Vector Current (CVC) relation, up to
numerically negligible IB corrections, an experimental determination of
the $I=1$ $K\bar{K}$-mode contribution to $\rho_{\rm EM}(s)$. In the region
where it is rather precise (up to $s=2.7556\ {\rm GeV}^2$, well above the
$\phi$ peak), the BaBar $\tau$ data can thus be used to provide a direct
determination of the $I=1$ $K\bar{K}$ contribution to $a_\mu^{\rm LO,HVP}$. 
For the contributions from the parts of the KNT19 and DHMZ exclusive-mode 
regions above $2.7556\ {\rm GeV}^2$, we employ the maximally conservative 
separation treatment of KNT19 results for the $K\bar{K}$ contributions to 
$R(s)$. The contributions from this region turn out to be much smaller 
than those from the region covered by the BaBar $\tau$ data. Note that KNT19 
$K\bar{K}$ input is used for the determination of this higher-$s$ $K\bar{K}$ 
exclusive-mode contribution for both the KNT19- and DHMZ-based versions of 
the analyses. The reason is that the full $K\bar{K}$ exclusive-mode data 
and covariances are publicly available only in the KNT19 case. Numerical 
details for the KNT19- and DHMZ-based analyses are provided below.

The exclusive-mode-region, $I=1$, $K\bar{K}\pi$-mode contributions to
$a_\mu^{\rm LO,HVP}$ are obtained using the Dalitz-plot-based $I=1/0$
separation of $K\bar{K}\pi$ cross-sections performed by
BaBar~\cite{BaBar:2007ceh}, an analysis made possible by the
observed saturation of the cross sections in the region of
interest for this paper by $KK^*$ contributions.

\subsection{pQCD and DV corrections in the inclusive region}
\label{inclregionsec}
For the pQCD expression used to represent $\rho_{\rm EM}^{I=1}(s)$ in the 
inclusive region, we employ the standard five-loop, $n_f=3$ pQCD 
result~\cite{Baikov:2008jh,Herzog:2017ohr} with PDG2020 input for 
$\alpha_s$~\cite{ParticleDataGroup:2020ssz}. $D=2$, perturbative
quark-mass-squared corrections are completely negligible. In the 
region from just below $\sqrt{s}=2$~GeV up to charm threshold, 
$n_f=3$ perturbative expectations for $R(s)$ are compatible within errors 
with the experimental determinations of BES~\cite{BES:2001ckj,BES:2009ejh} 
and KEDR~\cite{KEDR:2018hhr} (especially those of KEDR~\cite{KEDR:2018hhr}), 
but lie slightly below recent BESIII results~\cite{ BESIII:2021wib}. Small
residual DV contributions to $\rho_{\rm EM}(s)$ may thus be present even 
in this relatively large-$s$ region, making it important to estimate the 
impact of possible DV corrections to $\rho_{\rm EM}^{I=1}(s)$ in the 
inclusive region of our analyses as well. While the whole of the KNT19 
inclusive region lies in the region of agreement with BES and KEDR, the 
lower part of the DHMZ inclusive region extends to lower $s$, where the 
deviation of the perturbative expectation for $R(s)$ from the experimental 
sum-of-exclusive-mode-contributions determination is larger. An estimate 
of possible DV corrections to the pQCD approximation is thus of even 
more importance for the DHMZ-based analysis.

We investigate possible DV corrections using recent results for
DV contributions to the charged $I=1$ vector current spectral
function, $\rho_{\rm ud;V}(s)$, measured in hadronic $\tau$ decays.
$\rho_{\rm ud;V}(s)$ is related to the conventionally normalized EM
isovector spectral function, $\rho_{\rm EM}^{I=1}(s)$ by the CVC
relation, $\rho_{\rm EM}^{I=1}(s) ={\frac{1}{2}} \rho_{\rm ud;V}(s)$.
In Ref.~\cite{Boito:2020xli}, finite-energy sum rule (FESR) 
analyses of weighted integrals of a recently improved version of 
$\rho_{\rm ud;V}(s)$ were carried out using the large-$s$ ansatz
\begin{equation}
\left[\rho_{\rm ud;V}\right]_{\rm{DV}}(s)= \mbox{exp}
\left( -\delta_1 -\gamma_1 s\right)\, \sin\left( \alpha_1+\beta_1 s\right)\ ,
\label{lgencreggedvform}\end{equation}
for the DV contribution to $\rho_{ud;V}(s)$. As detailed in
Ref.~\cite{Boito:2017cnp}, this ansatz follows for massless quarks 
from large-$N_c$ and Regge arguments. The DV parameters $\delta_1$, 
$\gamma_1$, $\alpha_1$ and $\beta_1$, were obtained as part of the 
FESR fits. A range of different fits was considered, characterized by 
the choice of $s_{\rm min}$, the minimum $s$ for which the DV ansatz 
was to be employed. Of these, ten, with $s_{\rm min}$ lying between
$1.4251$ and $1.7256$ GeV$^2$, show excellent $p$-values, good 
stability of the DV parameter results between the different fits, 
and reasonably controlled DV parameter errors. With all these 
$s_{\rm min}$ lying below the onset of both the KNT19 and DHMZ 
inclusive regions, it is thus possible to consider integrated DV 
contributions to $a_\mu^{\rm lqc}$ in the inclusive region using 
any of these fits. We evaluate the DV contributions and associated
errors for each of these fits. We then take as the central value of
our estimate of the DV contribution the midpoint of the range 
spanned by these results and their errors, and as the uncertainty
on that estimate half of that range. Numerical details for the 
KNT19 and DHMZ cases are provided below.

\begin{boldmath}
\subsection{${{a_\mu^{\rm lqc}}}$ using KNT19 or DHMZ 
exclusive-mode input}
\label{knt19dhmzanalyses}
\end{boldmath}
We now turn to KNT19- and DHMZ-based determinations of $a_\mu^{\rm lqc}$,
neglecting small IB corrections, which will be dealt with in the next section.

We begin with the KNT19-based analysis. Reference~\cite{Boito:2022rkw}
provides the following results for the various KNT19-based contributions to
$a_\mu^{I=1}$.

The sum of all KNT19 $G$-parity-positive exclusive-mode contributions to
$a_\mu^{\rm LO, HVP}$ from the region $s\leq 3.7520\ {\rm GeV}^2$ is
\begin{equation}
\left[ a_\mu^{I=1}\right]_{G=+} =543.21(2.09) 
\times 10^{-10}\, .
\label{gposknt19amusum}\end{equation}
The breakdown, showing the modes which contribute and the KNT19
contribution from each, may be found in Table I of Ref.~\cite{Boito:2022rkw}.

The $I=1$ part of the $G$-parity-ambiguous $K\bar{K}$ contribution from
threshold to $3.7520\ {\rm GeV}^2$ is
\begin{equation}
\left[ a_\mu^{I=1}\right]_{K\bar{K}}=
0.85(9) \times 10^{-10}\, .
\label{knt19kkbarfromtau}\end{equation}

The $I=1$ part of the $G$-parity-ambiguous $K\bar{K}\pi$ contribution
from threshold to $3.7520\ {\rm GeV}^2$ is
\begin{equation}
\left[ a_\mu^{I=1}\right]_{K\bar{K}\pi} = 0.74(12)\times 10^{-10}\, .
\label{knt19kkbarpi}\end{equation}

The $I=1$ $K\bar{K}2\pi$ mode contribution is obtained by
first subtracting from the KNT19 $I=1+0$ total,
$\left[ a_\mu^{\rm LO,HVP}\right]_{K\bar{K}2\pi}=1.93(8)\times 10^{-10}$,
the purely $I=0$ component, $0.159(10)\times 10^{-10}$,
resulting from $e^+ e^-\rightarrow \phi\pi\pi$ with the $\phi$
subsequently decaying to $K\bar{K}$, and then applying the maximally 
conservative $50\pm 50\%$ assessment of the $I=1$ contribution to the
resulting still-ambiguous difference. The result is
\begin{equation}
\left[ a_\mu^{I=1}\right]_{K\bar{K}2\pi} = 0.89(89)\times 10^{-10}\, .
\label{knt19kkbar2pi}\end{equation}
The $I=0$ $\phi\pi\pi$ subtraction was evaluated using the
$e^+ e^- \rightarrow \phi\pi\pi$ cross sections reported
in Ref.~\cite{BaBar:2011btv}.

The sum of the total ($I=1+0$) contributions to $a_\mu^{\rm LO, HVP}$
from all remaining $G$-parity ambiguous KNT19 modes, as detailed in the 
Appendix of Ref.~\cite{Boito:2022rkw}, is $0.23(3)\times 10^{-10}$, 
leading to a maximally conservative estimate of the corresponding
$I=1$ contribution
\begin{equation}
\left[ a_\mu^{I=1}\right]_{\rm other} = 0.12(12)\times 10^{-10}\, .
\label{knt19residualgparambigcontrib}\end{equation}

The KNT19 inclusive region $I=1$ pQCD contribution is a factor $9/2$ times 
the corresponding strange-connected-plus-disconnected contribution 
reported in Ref.~\cite{Boito:2022rkw}, and hence
\begin{equation}
\left[ a_\mu^{I=1}\right]_{\rm pQCD} = 28.27(2)\times 10^{-10}\, .
\label{knt19pqcd}\end{equation}
The uncertainty reflects that on the input value of $\alpha_s$ and 
the estimated impact of $5$-loop truncation and is much smaller than 
the size of the estimated DV contribution, obtained as discussed
above,{\footnote{The DV contributions to $a_\mu^{I=1}$ obtained
from the ten fits noted above lie between $0.21(8)\times 10^{-10}$
and $0.26(12)\times 10^{-10}$, and hence cover the range from
$0.13\times 10^{-10}$ to $0.38\times 10^{-10}$. The central
value and error of the result quoted in Eq.~(\ref{knt19DV}) represent,
respectively, the midpoint and half the extent of this range.}}
\begin{equation}
\left[ a_\mu^{I=1}\right]_{\rm DV} = 0.26(12)\times 10^{-10}\, .
\label{knt19DV}\end{equation}
The total KNT19 inclusive region contribution, including this
estimate of the DV correction, is thus
\begin{equation}
\left[ a_\mu^{I=1}\right]_{\rm incl} = 28.53(26)\times 10^{-10}\, ,
\label{knt19incl}\end{equation}
where, to be conservative, we have assigned the full central
value of the estimated DV correction as an expanded uncertainty.

Adding the above contributions yields the following interim result, 
prior to applying IB corrections, for the $I=1$ contribution
\begin{equation}
a_\mu^{I=1} = 574.34(2.29)\times 10^{-10}
\label{knt19noIBtotalI1}\end{equation}
and hence the associated interim light-quark-connected result
\begin{equation}
a_\mu^{\rm lqc} = 638.16(2.55)\times 10^{-10}\, .
\label{knt19noIBtotallqc}\end{equation}
\vskip .15in
Turning now to the analogous DHMZ-based analysis, once more
relying heavily on results already detailed in Ref.~\cite{Boito:2022rkw},
we find the following results for the components of the DHMZ-based 
determination of $a_\mu^{I=1}$.

The sum of all DHMZ $G$-parity positive exclusive-mode contributions to
$a_\mu^{\rm LO, HVP}$ from the region $s\leq 3.24\ {\rm GeV}^2$ is
\begin{equation}
\left[ a_\mu^{I=1}\right]_{G=+}
=542.74(3.39)(1.12)_{\rm lin} 
\times 10^{-10}\, , \label{gposdhmzamusum}\end{equation}
where the first error is the quadrature sum of the statistical
and mode-specific, mode-to-mode-uncorrelated systematic errors
of Ref.~\cite{Davier:2019can}, and the second is the 100\%-correlated 
common systematic error. The subscript ``lin'' is a reminder of the 
fact that, as specified in Ref.~\cite{Davier:2019can}, this error is 
obtained by summing linearly the corresponding errors on the individual 
DHMZ exclusive-mode contributions.

The DHMZ contributions from $G$-parity-ambiguous exclusive modes 
in the region $s\leq 3.24\ {\rm GeV}^2$ are
\begin{eqnarray}
&&\left[ a_\mu^{I=1}\right]_{K\bar{K}}= 0.83(8) \times 10^{-10}\, ,
\label{dhmzkkbarfromtau}\\
&&\left[ a_\mu^{I=1}\right]_{K\bar{K}\pi} = 0.66(11)\times 10^{-10}\, ,
\label{dhmzkkbarpi}\\
&&\left[ a_\mu^{I=1}\right]_{K\bar{K}2\pi} = 0.37(37)\times 10^{-10}\, .
\label{dhmzkkbarpipi}\\
&&\left[ a_\mu^{I=1}\right]_{\rm other} = 0.00(1)\times 10^{-10}\, ,
\label{dhmzresidualgparambigcontrib}\end{eqnarray}
with all results, with the exception of that from the $K\bar{K}2\pi$ mode, 
given previously in Sec.~VI of Ref.~\cite{Boito:2022rkw}. The additional 
information needed to obtain the $K\bar{K}2\pi$ result is as follows. 
First, the DHMZ result for the total $I=1+0$ contribution to 
$a_\mu^{\rm LO, HVP}$ is $0.85(2)(5)(1)_{\rm lin}\times 10^{-10}$. 
Second, the $s\leq 3.24$ GeV$^2$, $I=0$ $\phi (\rightarrow K\bar{K})\pi\pi$ 
contribution implied by BaBar $e^+e^-\rightarrow \phi \pi\pi$
cross-sections~\cite{BaBar:2011btv} is $0.117(8)\times 10^{-10}$. 
The $G$-parity ambiguous $K\bar{K}2\pi$ remainder is thus 
$0.73(5)(1)_{\rm lin}\times 10^{-10}$ and the maximally conservative
assessment of the $I=1$ component thereof $0.37(37)\times 10^{-10}$,
where we have ignored all error components other than the strongly 
dominant maximally conservative separation uncertainty.

Finally, for the DHMZ inclusive region pQCD, DV and total contributions,
we find
\begin{eqnarray}
&&\left[ a_\mu^{I=1}\right]_{\rm pQCD} = 32.74(3)\times 10^{-10}\, ,
\label{dhmzpqcd}\\
&&\left[ a_\mu^{I=1}\right]_{\rm DV} = -0.19(31)\times 10^{-10}\, ,
\label{dhmzDV}\\
&&\left[ a_\mu^{I=1}\right]_{\rm incl} = 32.55(31)\times 10^{-10}\, .
\label{dhmzincl}\end{eqnarray}
We note that the uncertainty on the DV contribution in this case is 
significantly larger than that obtained for the KNT19 case 
above.{\footnote{The DV contributions to $a_\mu^{I=1}$ here
lie between $-0.34(16)\times 10^{-10}$ and $-0.04(16)\times 10^{-10}$, 
covering the range from $-0.50\times 10^{-10}$ to $0.12\times 10^{-10}$. 
The central value and error of the result quoted in Eq.~(\ref{dhmzDV}) 
represent, respectively, the midpoint and half the extent of this range.}}

Adding the above contributions yields the following interim DHMZ-based 
result 
\begin{equation}
a_\mu^{I=1} = 577.15(3.43)(1.12)_{\rm lin}\times 10^{-10}
\label{dhmznoIBtotalI1}\end{equation}
and hence the associated interim light-quark-connected result
\begin{equation}
a_\mu^{\rm lqc} = 641.28(3.81)(1.24)_{\rm lin}
\times 10^{-10}\, .
\label{dhmznoIBtotallqc}\end{equation}

The final step required to obtain the desired isospin-limit version,
$a_\mu^{\rm lqc;\, IL}$, of $a_\mu^{\rm lqc}$ is to apply EM and SIB 
corrections to the interim results (\ref{knt19noIBtotallqc}) and 
(\ref{dhmznoIBtotallqc}). This step is discussed in the next section.

\vspace{0.8cm}
\section{EM and SIB corrections}\label{IBcorrns}
In this section we consider EM and SIB corrections to the results above.
In assessing EM corrections, where it is not feasible to carry out
a dispersive determination, we take advantage of the results of
the recent BMW lattice study~\cite{Borsanyi:2020mff}, which provides the 
first determination of all EM contributions to $a_\mu^{\rm LO, HVP}$. This 
(unavoidable) introduction of lattice EM input into an otherwise purely 
dispersive determination of $a_\mu^{\rm lqc;\, IL}$ turns out to constitute 
a rather minor ``deviation,'' since the lattice result for the EM 
correction is, in fact, rather small.

The separation of the IB sum of EM and SIB contributions into 
separate EM and SIB parts is, as is well known, ambiguous at 
$O\left( \alpha_{\rm EM}(m_d+m_u)\right)$.{\footnote{For an expanded 
discussion, see, {\it e.g.}, Sections 3.1.1 and 3.1.2 of 
Ref.~\cite{FlavourLatticeAveragingGroup:2019iem}.}}
The separation scheme used by BMW~\cite{Borsanyi:2020mff} in determining 
the EM corrections we employ below is defined such that the EM contributions 
to the masses of the purely connected neutral pseudoscalar mesons are 
zero, {\it i.e.}, such that all such EM contributions are absorbed 
into the definitions of the quark masses. It is numerically very 
similar to the widely used GRS scheme~\cite{Gasser:2003hk}.
By using BMW EM results, we are working in the BMW separation 
scheme.{\footnote{This is, for our purposes, a somewhat academic 
point since $m_d+m_u$ is only a factor of $\sim 3$ greater than 
$m_d-m_u$, making the $O\left( \alpha_{\rm EM}(m_d+m_u)\right)$ 
separation ambiguity comparable in size to contributions of
$O\left( \alpha_{\rm EM}(m_d-m_u)\right)$ which, being second order
in IB, we are neglecting throughout.}}

To perform the desired EM and SIB corrections, one needs to identify and 
subtract EM and SIB contributions present in the experimental versions of 
the nominal $I=1$ contribution $a_\mu^{I=1}$ determined in the previous
section. These are of two types: those present in the physical $a_\mu^{I=1}$ 
contribution itself, and those associated with nominally $G$-parity 
positive contributions which are actually part of the mixed-isospin 
contribution, $a_\mu^{\rm{MI}}$, and which hence ``contaminate'' the nominal
$a_\mu^{I=1}$ results obtained above. The correction for the mixed-isospin 
contamination cannot be done inclusively since $\rho_{\rm EM}^{\rm MI}(s)$ 
receives contributions from both nominally $G$-parity positive and 
nominally $G$-parity negative exclusive modes, with, {\it e.g.}, 
$\rho$-$\omega$ mixing inducing both mixed-isospin $\pi^+\pi^-$ and 
mixed-isospin $\pi^+\pi^-\pi^0$ contributions, via, respectively, the 
processes $e^+e^-\rightarrow\omega\rightarrow\rho\rightarrow\pi^+\pi^-$ and
$e^+e^-\rightarrow\rho\rightarrow\omega\rightarrow\pi^+\pi^-\pi^0$.

To first order in IB, SIB contributions to $\hat{\Pi}_{\rm EM}(Q^2)$ and 
$\rho_{\rm EM}(s)$ occur only in the mixed-isospin parts, while EM 
contributions are present in all of the $I=1$, $I=0$ and mixed-isospin
components. It follows that, to this order, the correction required to 
convert the  physical version of $a_\mu^{33}=a_\mu^{I=1}$ to the 
corresponding isospin-limit version is purely EM in nature. We denote 
the associated contribution to $a_\mu^{\rm lqc}$, to be subtracted 
from the nominal $a_\mu^{\rm lqc}$ obtained in the previous section, by 
$\delta_{\rm EM} a_\mu^{\rm lqc}$. There is no need for a breakdown of 
this correction into components associated with individual exclusive modes,
and we take as input for this contribution the inclusive lattice 
result quoted in Ref.~\cite{Borsanyi:2020mff},
\begin{equation}
\delta_{\rm EM} a_\mu^{\rm lqc}\, =\, -0.93(34)(47)\times 10^{-10}\, ,
\label{bmwamuieq1emcorrn}\end{equation}
where the first error is statistical and the second systematic.
As noted above, the associated correction represents a very small 
fraction of the nominal $a_\mu^{\rm lqc}$ results above.

We now address the mixed-isospin-contamination correction. Evaluating
this correction requires identifying and subtracting mixed-isospin 
contaminations present in each of the individual exclusive-mode 
contributions summed to obtain the nominal $a_\mu^{\rm lqc}$ results 
of the previous section. In contrast to $a_\mu^{I=1}$, which, to first 
order in IB, receives no SIB contribution, both EM and SIB contributions
are present in $a_\mu^{\rm MI}$, with SIB expected to dominate. In what
follows, we rely on experimental input to quantify what should be the
dominant exclusive-mode correction and provide conservative bounds on 
the remaining sub-dominant contributions. In relying on experimental
input, the results for the mixed-isospin corrections are, of course,
those for the sum of EM and SIB effects. 

As is well known, the strong low-$s$ enhancement produced by the dispersive
weight $\hat{K}(s)/s^2$ is such that the dispersive determination of 
$a_\mu^{\rm{LO,HVP}}$ is dominated by contributions from the region of the 
lowest-lying (especially $\rho$ and $\omega$) resonances. A similar low-$s$,
resonance-region dominance is expected for $a_\mu^{\rm MI}$, doubly so since
IB contributions in this region are subject to enhancements generated by the 
impact of the very small $\rho$-$\omega$ mass difference on contributions 
induced by $\rho$-$\omega$ mixing. In this region, the mixed-isospin 
spectral contribution, $\rho_{\rm{EM}}^{\rm{MI}}(s)$, will appear 
essentially entirely in the $2\pi$ and $3\pi$ exclusive modes. Prior to
implementing the mixed-isospin correction, the IB $2\pi$ and $3\pi$ components 
appear, respectively, in the nominal $I=1$ and $I=0$ sums, and hence represent
mixed-isospin contaminations of those sums. In this study, we are interested 
only in carrying out the mixed-isospin correction for the nominal $I=1$ sum, 
and hence focus on the IB contribution to the $2\pi$ distribution.
Note that more than 90\% of the full exclusive-mode-region contribution 
to $a_\mu^{\rm lqc}$, in fact, comes from $\pi\pi$ contributions in the 
region below $s=1\ {\rm GeV}^2$.

The presence of the obvious $\rho$-$\omega$ interference ``shoulder''
in the $e^+e^-\rightarrow\pi^+\pi^-$ cross section makes possible an 
experimental determination of the IB $\rho$-$\omega$-region contribution 
from the $\pi\pi$ exclusive-mode. To first order in IB, the associated
low-$s$ ($s<1\ {\rm GeV}^2$) contribution to $a_\mu^{\rm{LO,HVP}}$ lies 
entirely in $a_\mu^{\rm{MI}}$. This contribution, which should strongly 
dominate $\left[a_\mu^{\rm{MI}}\right]_{\pi\pi}$, has recently been 
determined using the results of a fit to the $e^+e^-\rightarrow\pi^+\pi^-$ 
cross sections based on a dispersively constrained representation of the 
timelike $\pi$ form factor incorporating the IB $\rho$-$\omega$ 
interference effect~\cite{Colangelo:2022prz}. The use of rigorous 
dispersive constraints turns out to produce a rather tightly constrained 
result,
\begin{equation}
\left[a_\mu^{\rm{MI}}\right]_{\pi\pi}=3.68(14)(10)\times 10^{-10}\, ,
\label{chks22MI2pi}\end{equation}
where the first error is the fit uncertainty and the second the
combination of systematic uncertainties.
The associated mixed-isospin $\pi\pi$ contamination of the nominal
$a_\mu^{\rm lqc}$ obtained in the last section, which we denote
$\left[\delta_{\rm MI} a_\mu^{\rm lqc}\right]_{\pi\pi}$, is thus 
\begin{equation}
\left[\delta_{\rm MI} a_\mu^{\rm lqc}\right]_{\pi\pi}=4.09(16)(11)
\times 10^{-10}\, .
\label{mi2pilqccontamcontrib}\end{equation}
This contribution must be subtracted from the nominal $a_\mu^{I=1}$
({\it i.e.} $a_\mu^{\rm lqc}$) results of the previous section. 

Note that, in spite of the strong, narrow resonance enhancement, 
the IB contribution in Eq.~(\ref{chks22MI2pi}) represents only 
$0.7\%$ of the full exclusive-mode-region $\pi\pi$ contribution 
$\left[a_\mu^{\rm{LO, HVP}}\right]_{\pi\pi}$. Since contributions to
the nominal $a_\mu^{I=1}$ total from exclusive modes other than $\pi\pi$ 
are dominated by regions in $s$ for which no analogous narrow interfering 
resonance enhancements are possible, it should be extremely conservative 
to assume the magnitudes of contributions to $a_\mu^{\rm MI}$ from all 
non-$\pi\pi$ exclusive modes are also less than $\sim 1\%$ of the 
corresponding nominal $a_\mu^{I=1}$ exclusive-region contributions. The 
non-$\pi\pi$ exclusive-mode-region contributions to the nominal $I=1$
sum $a_\mu^{I=1}$ total $41.6\times 10^{-10}$ in the KNT19 case and 
$36.7\times 10^{-10}$ in the DHMZ case. We thus expect the sum of 
exclusive-mode-region contributions to $a_\mu^{\rm MI}$ from all 
exclusive modes in that sum other than $\pi\pi$ to be less than 
$0.42\times 10^{-10}$ and $0.38\times 10^{-10}$ in 
magnitude, respectively, for the KNT19 case and DHMZ cases. Unsurprisingly, 
these bounds are much smaller than the accurately determined $\pi\pi$ 
contribution given in Eq.~(\ref{chks22MI2pi}). We thus take, as our 
estimate of the full mixed-isospin contamination present in the nominal 
$a_\mu^{I=1}$ sums of the previous section, the result of 
Eq.~(\ref{chks22MI2pi}), adding the non-$\pi\pi$ exclusive-mode bounds 
as additional systematic uncertainties. 

Adding the results of Eqs.~(\ref{bmwamuieq1emcorrn}) and
(\ref{mi2pilqccontamcontrib}), we find, for the sum of EM+SIB contributions,
$\delta_{\rm EM+SIB} a_\mu^{\rm lqc}\equiv \delta_{\rm EM} a_\mu^{\rm lqc}
+\left[\delta_{\rm MI} a_\mu^{\rm lqc}\right]_{\pi\pi}$, to be subtracted 
from the nominal light-quark-connected results, Eqs.~(\ref{knt19noIBtotallqc})
and (\ref{dhmznoIBtotallqc}), to convert to the corresponding isospin-limit 
values, $a_\mu^{\rm lqc;\, IL}$, the results 
$\delta_{\rm EM+SIB}a_\mu^{\rm lqc} =3.16(37)(48)(46)\times 10^{-10}$ 
and $3.16(37)(48)(41)\times 10^{-10}$ for the KNT19 and DHMZ cases, 
respectively, where the first errors are statistical, the second are
systematic, and the third are the uncertainties estimated above
for missing mixed-isospin contributions from non-$\pi\pi$ exclusive modes. 
Combining, for simplicity of presentation, all errors in quadrature, we 
obtain the following final results for $a_\mu^{\rm lqc, IL}$:
\begin{eqnarray}
&& a_\mu^{\rm lqc;\, IL}=635.0(2.7)\times 10^{-10} \quad {\rm(KNT19)}
\label{knt19ILamulqc}\\
&& a_\mu^{\rm lqc;\, IL}=638.1(4.1)\times 10^{-10} \quad {\rm(DHMZ)}\, .
\label{dhmzILamulzc}\end{eqnarray}

\section{Conclusions and discussion}\label{conclusions}
\begin{figure}[t]
\begin{center}
\includegraphics[width=.8\textwidth,angle=0]
{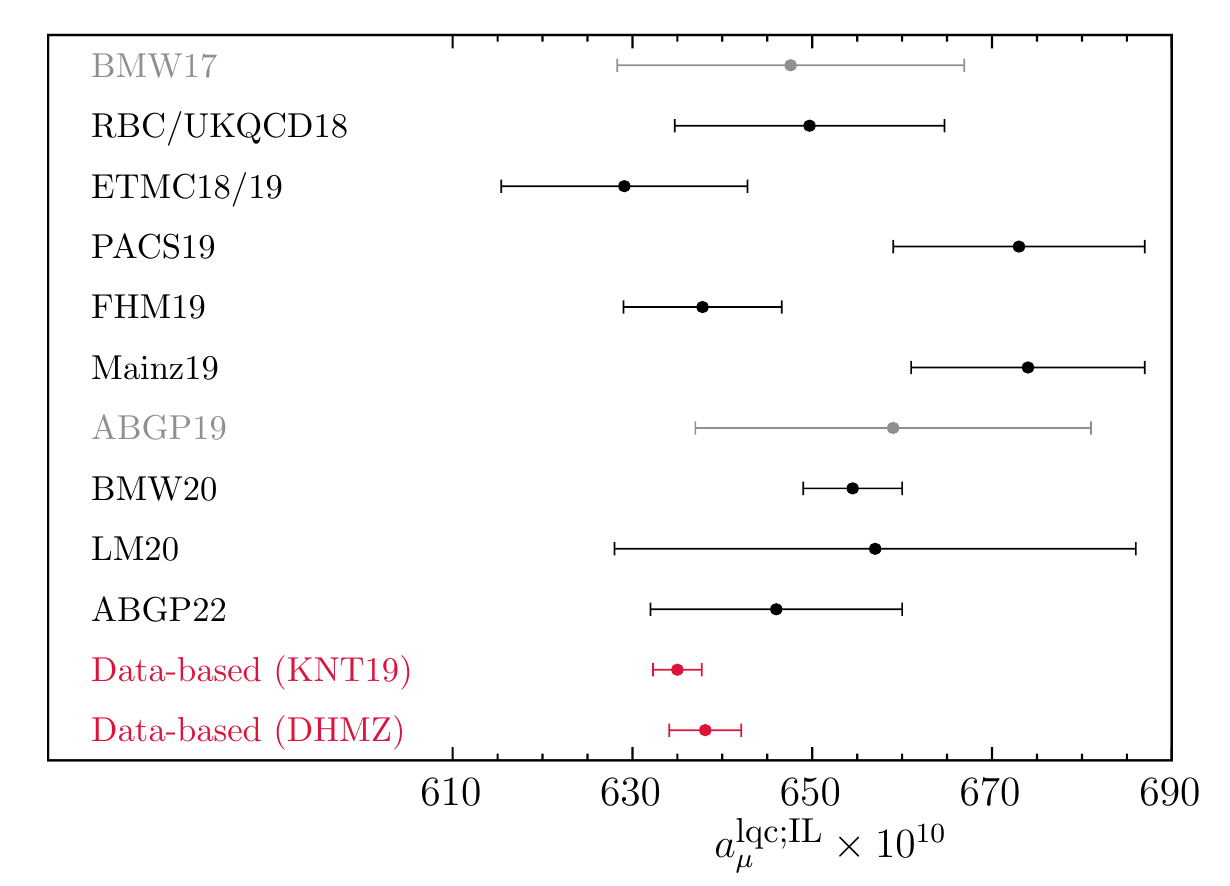}
\caption{\label{displattillqccomp} {\it Comparison of lattice determinations 
and our dispersive results for $a_\mu^{\rm lqc,\, IL}$. Lattice results
superseded by those from later publications by the same collaboration
are plotted in grey. See Table~\ref{tab1}
for the corresponding numerical values and references.}}
\end{center}
\end{figure}

We have shown how recent dispersive results for exclusive-mode 
contributions to $a_\mu^{\rm LO, HVP}$ can be used to provide a 
determination of the corresponding isospin-limit, light-quark-connected 
contribution, $a_\mu^{\rm lqc;\, IL}$, the precision of which turns out to 
be of order $0.5\%$. The determination employs lattice input for a small, 
$0.15\%$, EM correction, but is otherwise purely dispersive. The result, 
of course, depends on the choice of exclusive-mode input, and small 
differences in the KNT19 and DHMZ assessments of individual exclusive-mode
contributions lead to an associated $\sim 0.5\%$ difference between the
$a_\mu^{\rm lqc;\, IL}$ results obtained using KNT19 and DHMZ input,
given in Eqs.~(\ref{knt19ILamulqc}) and (\ref{dhmzILamulzc}), 
respectively. This difference is similar in size to the errors on the 
individual KNT19- and DHMZ-based determinations, and sufficiently small 
to allow meaningful conclusions to be drawn from a comparison of our 
dispersive results to those of recent lattice analyses. This comparison 
is summarized in Table~\ref{tab1} and Figure~\ref{displattillqccomp}.
Our dispersive results 
lie lower than the majority of central lattice values, though some 
variability, at the roughly $2\sigma$ level, remains in the lattice 
results. Among the lattice results, that of Ref.~\cite{Borsanyi:2020mff} 
(BMW 2020) has, at present, by far the smallest error, and would strongly 
dominate any putative lattice average. Our KNT19- and DHMZ-based dispersive 
results are in $3.2$ and $2.4\, \sigma$ tension, respectively, with the BMW 
2020 result. Other lattice results, from multiple groups, with similar 
or smaller errors, are anticipated in the near future, 
and our dispersive results provide a useful comparison target for 
such future lattice determinations.

\begin{table}[t]
\begin{center}
\begin{tabular}{ll}
\hline
$a_\mu^{\rm lqc;\, IL}\times 10^{10}$&Reference\\
\hline
647.6(19.3)&BMW 2017~\cite{Budapest-Marseille-Wuppertal:2017okr}\\
649.7(15.0)&RBC/UKQCD 2018~\cite{RBC:2018dos}\\
629.1(13.7)&ETMC 2018/2019~\cite{Giusti:2018mdh,Giusti:2019hkz}\\
673(14)&PACS 2019~\cite{Shintani:2019wai}\\
637.8(8.8)&FHM 2019~\cite{FermilabLattice:2019ugu}\\
674(13)&Mainz 2019~\cite{Gerardin:2019rua}\\
659(22)&ABGP 2019~\cite{Aubin:2019usy}\\
654.5(5.5)&BMW 2020~\cite{Borsanyi:2020mff}\\
657(29)&LM 2020~\cite{Lehner:2020crt}\\
646(14)&ABGP 2022~\cite{Aubin:2022hgm}\\
\hline
635.0(2.7)&This work (KNT19-based)\\
638.1(4.0)&This work (DHMZ-based)\\
\hline
\end{tabular}
\caption{\label{tab1} {\it Comparison of our dispersive results with 
recent lattice results for $a_\mu^{\rm lqc;IL}$. The latter
are listed above the internal horizontal line, the former below it.}} 
\end{center}
\end{table}

The analysis strategy employed above, though applied there only to the 
dispersive determination of $a_\mu^{\rm lqc;\, IL}$, is readily adapted 
to determinations of other quantities of interest also having a dispersive 
representation. One well-known example is the standard intermediate window 
quantity, $a_\mu^W$, introduced by RBC/UKQCD~\cite{RBC:2018dos}, and 
constructed, by design, to be rather precisely determinable on the lattice. 
The isospin-limit, light-quark-connected component of $a_\mu^W$, 
$a_\mu^{\rm W,lqc;\, IL}$, has now been determined by a number of lattice 
groups~\cite{RBC:2018dos,Aubin:2019usy,Borsanyi:2020mff,Lehner:2020crt,Giusti:2021dvd,Wang:2022lkq,Aubin:2022hgm,Ce:2022kxy,Alexandrou:2022amy,RBC22,FHM22},
with updates of earlier ABGP~\cite{Aubin:2019usy}, ETMc~\cite{Giusti:2021dvd} 
and RBC/UKQCD~\cite{RBC:2018dos} results, reported in
Refs.~\cite{Aubin:2022hgm,Alexandrou:2022amy,RBC22}, bringing results from 
all groups into excellent agreement. These results are also found to lie 
significantly higher than alternate mixed ``R-ratio+lattice'' estimates 
obtained by subtracting from R-ratio-based dispersive determinations of 
$a_\mu^W$ contributions for all non-light-quark-connected components 
evaluated on the lattice.
Of course, in view of other signs of tension between lattice and 
dispersive results, one would prefer to compare the rather precise 
lattice results with a dispersive, rather than mixed lattice-dispersive 
expectation. This is not currently possible because no purely dispersive 
$a_\mu^{\rm W,lqc;\, IL}$ determination exists. 

Weighted exclusive-mode and inclusive-region integrals are, however, equally 
easy to evaluate for any dispersive weight as they are for the weight 
$\hat{K}(s)/s^2$ which enters the determinations of the exclusive-mode 
$a_\mu^{\rm LO, HVP}$ contributions of Refs.~\cite{Davier:2019can,KNT19}, 
provided, that is, the relevant exclusive-mode distributions, 
$\left[\rho_{\rm EM}(s)\right]_X$, required  to perform these re-weighted 
integrals, are publicly available. This information {\it is} available for 
the distributions underlying the KNT19 exclusive-mode results of 
Ref.~\cite{KNT19}, and KNT19-based dispersive determinations of, not 
just $a_\mu^{\rm W, lqc;\, IL}$, but also other dispersive integral 
quantities, involving different dispersive weights, are thus also 
possible using the analysis strategy above. 

Quantities of this type likely to be of interest for future investigation 
include both those naturally formulated in terms of their Euclidean-time 
($t$) weightings and those naturally formulated in terms of their 
dispersive $s$-weightings. Examples of the former include the 
additional intermediate window quantities of Ref.~\cite{RBC:2018dos}, 
the one-sided window quantities of Ref.~\cite{FermilabLattice:2022izv}, 
the linear-combinations-of-Euclidean-window quantities of 
Ref.~\cite{Colangelo:2022vok}, and the window quantity, $a_\mu^{W2}$, 
introduced in Ref.~\cite{Aubin:2022hgm}, designed to more strongly weight 
higher-$t$ lattice contributions and improve the reliability of ChPT-based 
estimates of lattice finite-volume effects. Examples of the latter are 
dispersive integrals involving $s$-dependent weights of the type introduced 
in Ref.~\cite{Boito:2022njs}, designed to emphasize contributions from more 
limited regions in $s$ and potentially help in obtaining a more detailed 
understanding of the source of the current dispersive-lattice tensions. 
Since the sum of isospin-limit strange-quark-connected and full-disconnected 
contributions to $\rho_{\rm EM}(s)$ also has a representation as an 
appropriately weighted difference of nominally $I=0$ and nominally 
$I=1$ exclusive-mode contributions~\cite{Boito:2022rkw}, the current analysis 
strategy can be employed to determine isospin-limit versions of not just 
light-quark-connected, but also strange-connected-plus-full-disconnected, 
contributions to all such window quantities. We plan to implement such
dispersive determinations of these various window quantities and will 
report on the results of this work in a future paper.

\vspace{3ex}
\noindent {\bf Acknowledgments}
This material is based upon work supported by the U.S. Department 
of Energy, Office of Science, Office of Basic Energy Sciences Energy 
Frontier Research Centers program under Award Number DE-SC-0013682.
DB's work was supported by the S\~ao Paulo Research Foundation (FAPESP) 
Grant No. 2021/06756-6 and by CNPq Grant No. 308979/2021-4. The work 
of KM is supported by a grant from the Natural Sciences and Engineering 
Research Council of Canada. SP is supported by the Spanish Ministry 
of Science, Innovation and Universities (project 
PID2020-112965GB-I00/AEI/10.13039/501100011033) and by Grant 2017 SGR 
1069. IFAE is partially funded by the CERCA program of the Generalitat 
de Catalunya.
\vspace{3ex}


\end{document}